\def\keV{{\rm keV}}

\def\rmp{Rev. Mod. Phys.}

\documentclass[preprint2]{aastex}
\usepackage{epsfig}
\usepackage{graphicx}
\usepackage{natbib}
\DeclareGraphicsExtensions{.epsi,.ps,.eps}
\begin{document}
\shorttitle{FRAMEWORK FOR AGN HEATING}
\shortauthors{VOIT \& DONAHUE}
\slugcomment{\apj, in press}
\title{An Observationally Motivated Framework for AGN Heating of Cluster Cores}
\author{G. Mark Voit\altaffilmark{1}$^,$\altaffilmark{2} \& 
              Megan Donahue\altaffilmark{1}$^,$\altaffilmark{3} 
         } 
\altaffiltext{1}{Department of Physics and Astronomy,
                 Michigan State University, 
                 East Lansing, MI  48824}
\altaffiltext{2}{voit@pa.msu.edu}
\altaffiltext{3}{donahue@pa.msu.edu}

\begin{abstract}
The cooling-flow problem is a long-standing puzzle that has received considerable recent attention, in part because the mechanism that quenches cooling flows in galaxy clusters is likely to be the same mechanism that sharply truncates the high end of the galaxy luminosity function.  Most of the recent models for halting cooling in clusters have focused on AGN heating, but the actual heating mechanism has remained mysterious.  Here we present a framework for AGN heating derived from a {\em Chandra} survey of gas entropy profiles within cluster cores.  This set of observations strongly suggests that the inner parts of cluster cores are shock-heated every $\sim 10^8$ years by intermittent AGN outbursts, driven by a kinetic power output of $\sim 10^{45} \, {\rm erg \, s^{-1}}$ and lasting at least $10^7$ years.  Beyond $\sim 30$~kpc these shocks decay to sound waves, releasing buoyant bubbles that heat the core's outer parts.  Between heating episodes, cooling causes the core to relax toward an asymptotic pure-cooling profile.  The density distribution in this asymptotic profile is sufficiently peaked that the AGN shock does not cause a core entropy inversion, allowing the cluster core to retain a strong iron abundance gradient, as observed.
\end{abstract}


\setcounter{footnote}{0}

\section{Introduction}

The gaseous cores of galaxy clusters have defied interpretation for almost three decades.  Initially, gas in the cores of clusters was thought to cool, condense, and flow toward the center, as long as the time required for the central gas to radiate away its thermal energy was less than the age of the universe \citep{fn77, cb77, mb78}.  Most galaxy clusters satisfy this condition, and those that do have traditionally been called cooling-flow clusters. Early estimates of their mass accretion rates ranged as high as $10^2 - 10^3 \, M_\odot \, {\rm yr}$ \citep{Fabian94}. The problem with this interpretation of the data was that the mass sink for all this supposedly cooling and condensing gas has never been found \citep{bd94,mj94,Odea94,opk98,
vd95,dv04}.

Spectroscopic X-ray observations have now established that the rates of cooling and condensation in galaxy clusters are much lower than those originally inferred from the imaging observations \citep{pet_etal_01,pet_etal_03, Tamura01}.  There is little evidence for emission lines from gas cooling much below about half the ambient temperature of the hot gas filling the cluster.  The cooling process seems to have stopped in its tracks, implying that a compensating heat source resupplies the energy radiated by the gas in cluster cores.

Supermassive black holes at the centers of galaxy clusters are an attractive candidate for supplying that heat energy \citep[e.g.,][]{bt95}.  Clusters suspected of harboring cooling flows are statistically more likely to contain radio emission indicative of recent black-hole activity \citep[e.g.,][]{Burns90}, and that radio emission often coincides with cavities in the X-ray emitting plasma, clearly showing that outflows from the nuclei of galaxies are interacting with the core gas in clusters \citep[e.g.,][]{Mcnamara00,
Fabian00}.  Measuring the work needed to inflate these cavities reveals that each outburst of energy from an active galactic nucleus (AGN) introduces $\sim 10^{58} - 10^{61} \, {\rm erg}$ into the intracluster medium \citep{brmwn04,McNamara05}, which would be sufficient to stop the cooling if such outbursts occurred every $\sim 10^8$ years or so, supplying a time-averaged energy input rate $\sim 10^{43} - 10^{45} \, {\rm erg \, s^{-1}}$ \citep{Churazov02, David01}. 

While AGN heating is certainly plausible, it has not been clear how the outburst energy gets transfered to the intracluster gas \citep[cf.][]{bblp02}.  Suggestions that the AGN outflow heats the ambient gas by driving a powerful shock \citep[e.g.,][]{swdm01} were dismissed when early {\em Chandra} observations showed cool rims rather than hot ones around the X-ray cavities, motivating investigations into gentler heating mechanisms \citep{Churazov00,Churazov01, Begelman01,rb02, qbb01, rhb02, bk02, DallaVec04}.  The cool rims now seem to consist of low-entropy gas transported to larger radii by the buoyantly rising cavities, but there are other reasons  why gentle heating mechanisms have been preferred.   One is that clusters with short central cooling times have iron abundances that strongly decrease with radius \citep[e.g.][]{delm04}.  An impulsive injection of energy at the center of the cluster would appear to destroy such a gradient by causing the cluster core to convect, thereby mixing the iron more uniformly \citep[but see][]{obbs04}.  Another is that the abrupt jumps in gas pressure expected of shocks were not obvious in much of the early {\rm Chandra} data;  however, detecting a localized region of hot gas can be quite difficult if emission from lower-temperature gas is present along the same line of sight \citep[e.g.,][]{mrmt04}.

Some recent X-ray observations are suggesting that the shock-heating hypothesis may have been prematurely discarded.  Deep {\em Chandra} images of Hydra A \citep{Nulsen05} and MS0735+7421 \citep{McNamara05} have revealed that AGN activity can drive shock fronts to distances $\sim 200-300$~kpc from the center, indicating that a kinetic power output $\sim 10^{45} - 10^{46} \, {\rm erg \, s^{-1}}$ has persisted for $\sim 10^8$ years in those clusters.  Cases like these may be rare, but objects that currently show radio emission indeed seem systematically different from ones that do not.  \citet{chb05} have reported that groups of galaxies with radio emission have systematically lower X-ray luminosities and higher X-ray temperatures than those without radio emission.  They interpret this difference as transient spikes in temperature because the ratio of optical to X-ray luminosity shows no such systematic difference.  

Another clue comes from {\em Chandra} observations of two radio-quiet clusters by \citet{don05_rq}. Both clusters were classified as strong cooling flows by \citep{Peres98}, but their central cooling times ($\sim 10^9$ years) are much longer than the central cooling times of cooling-flow clusters with current radio activity ($\sim 10^8$ years).  Either the gas in these clusters has never cooled and condensed, obviating the need for feedback, or they experienced such a dramatic feedback episode sometime in the past that they have not required any additional feedback for $\sim 10^9$ years.  The fact that signs of AGN feedback appear only in those cooling-flow clusters that currently require feedback strengthens the case for AGN heating as the mechanism that limits cooling.

Here we revisit shock heating in the context of a {\em Chandra} sample of classic cooling-flow clusters \citep{don05_cc}.  Profiles of gas entropy in the cores of these clusters show a striking consistency that stands as a strong constraint on any model purporting to explain the cooling-flow problem.  Section 2 assesses the clues these profiles hold about the cooling and feedback mechanisms in cluster cores, while Section 3 outlines a framework for AGN heat input consistent with those clues.  The picture we arrive at from analyzing these entropy profiles agrees in many respects with the AGN-heating models inspired by the X-ray cavities and is aligned with the suggestion of \citet[][and references within]{Binney05} that episodic outflows from AGNs quench cooling and condensation in clusters, limit the maximum luminosity of galaxies, and regulate the growth of black holes at their centers.  We show that outbursts of $\sim 10^{45} \, {\rm erg \, s^{-1}}$ occurring on a $\sim 10^8$~year timescale are sufficient to maintain the observed core entropy profiles.  However, unusually powerful outbursts like those seen in Hydra~A and MS0435+7241, which are rarer and longer-lasting than those needed to maintain the observed profiles, may also play a role in extending the time between outbursts and in elevating intracluster entropy beyond the cores of clusters.  Section 4 summarizes our results.

\section{Core Entropy Profiles}
\label{sec-coreprofs}

We focus here on the entropy content of intracluster gas rather than on temperature and gas density individually because entropy offers more direct insights into the processes that add and remove thermal energy from the gas.  Adding heat energy to a cluster does not simply raise the gas temperature.  Instead, the heat input causes the cluster's gas to expand within its potential well, and the temperature it achieves after relaxing to equilibrium depends primarily on its entropy distribution and the shape of that potential well \citep[e.g.,][]{vbbb02, Voit_RMP}.  In this paper we will quantify entropy with the adiabatic constant $K \equiv kT n_e^{-2/3}$, because this quantity provides the most direct link between X-ray observations and entropy content.

\begin{figure}[t]
\includegraphics[width = 3in,trim = 1.0in 1.3in 1.0in 1.0in, clip]{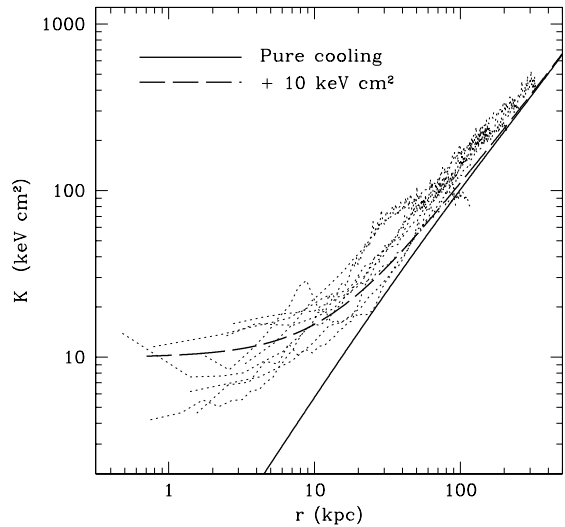}
\caption{ \footnotesize
Core entropy profiles of nine classic cooling-flow clusters from \citet{don05_cc}.  The dotted lines show the gas entropy ($K = k T n_e^{-2/3}$) profiles derived from the azimuthally averaged and deprojected density and temperature profiles of these clusters as a function of radius.  The profiles look strikingly similar considering that the cluster temperatures in this sample range from 2.2~keV to 7.4~keV.  Similarity in the entropy levels of the outer regions is consistent with the $T^{2/3}$ scaling of the outer entropy profiles of clusters seen by \citet{psf03}.  At larger radii the profiles seem to asymptotically approach the solid line showing the expected entropy profile for a 5 keV cluster model from \citet{vbbb02} in which radiative cooling acts without triggering feedback.  Simply adding a constant $10 \, {\rm keV \, cm^2}$ to the asymptotic model gives the thick dashed line, which is a good representation of the observed profiles within about 50~kpc (assuming $H_0 = 70 \, {\rm km \, s^{-1} \, Mpc^{-1}}$).
\label{profs+mod+10}}
\end{figure}

Figure \ref{profs+mod+10} shows nine core entropy profiles of cooling-flow clusters taken by \citet{don05_cc} from the {\em Chandra} archive: 2A0335+096, Abell~133, Abell~262, Abell~496, Abell~1795, Abell~2029, Abell~2052, Hydra A, and PKS 0745+191.  The data were azimuthally averaged in annuli centered on the cluster core and then deprojected, giving $n_e$ and $T$ as functions of radius, which were then combined into $K(r)$ profiles.  The similarity among the profiles immediately suggests that some sort of quasi-steady process stabilizes the intracluster gas.  There are no large entropy inversions outside of 10~kpc, and the apparent entropy inversion seen at $\approx 9$~kpc in one of the clusters (Abell~2052) arises primarily from non-axisymmetric structure owing to its X-ray cavities.  

Also notable is the fact that entropy at a given physical radius is quite similar from cluster to cluster, even though the clusters range in temperature from 2.2 keV to 7.4 keV.  That form of similarity can be understood if the entropy profiles of clusters scale with the characteristic halo temperature $T_\phi$ according to $K \propto T_\phi^{2/3}$, as suggested by the data of \citet{psf03}.  \citet{don05_cc} find that the dependence of entropy on radius in these clusters is $K(r) \propto r^\alpha$ with $\alpha \approx 0.9 - 1.3$.  Assuming that $\alpha = 1.1$ and that the virial radius of a cluster's halo is $ \propto T_\phi^{1/2}$, one finds that $K(r) \propto T_\phi^{2/3} (r/T_\phi^{1/2})^{1.1} \propto T_\phi^{0.12} r^{1.1}$.  In other words, the $K(r)$ profiles depend only weakly on temperature, as long as the entropy at a given fraction of the virial radius is $\propto T_\phi^{2/3}$.  It is not yet clear whether this altered entropy scaling stems mostly from cooling and supernova feedback during the epoch of galaxy formation or whether it requires additional AGN heating at lower redshifts \citep[e.g.,][]{bv05}

The thick solid line in Figure \ref{profs+mod+10} shows the core entropy profile one would expect for a 5 keV cluster in which radiative cooling operates without any feedback, according to \citet{vbbb02}.  Their model started with the baseline entropy profile expected from gravitational structure formation alone and then allowed that gas to cool for a Hubble time within a Navarro-Frenk-White potential \citep{nfw97}.  The inner temperature profiles predicted by that pure cooling model (the $\hat{K}_{\rm R}$ model in the lingo of \citet{vbbb02}) were shown to be quite similar to those of clusters suspected to harbor strong cooling flows, suggesting that cooling indeed plays an important role in determining their core entropy structure \citep[see also][]{McCarthy04}.

\citet{don05_cc} show that cooling-flow clusters do not actually relax to the pure-cooling profile.  Instead, they appear to asymptotically approach it at $\sim 100$~kpc.  Meanwhile, the inner entropy profiles of these clusters flatten out at the $\sim 10 \, {\rm keV \, cm^2}$ level, a clue to the form of heat input that hinders cooling and condensation.  Assuming pure free-free cooling, one can write a simple relation between entropy, cooling time, and gas temperature:
\begin{equation}
  t_c \approx 10^8 \, {\rm yr} \; \left( \frac {K} {10 \, {\rm keV \, cm^2}} \right)^{3/2}  
                                                       \left( \frac{kT} {5 \, {\keV}} \right)^{-1} \; \; .
  \label{eq-ctime}
\end{equation}
The central flattening of these clusters' entropy profiles therefore suggests that they will not require feedback to halt cooling until $\sim 10^8$ years from now.  In that respect, the clusters in the \citet{don05_cc} sample follow a trend first noticed in Hydra~A by \citet{David01}.  Heating would therefore seem to be episodic \citep{kb03}.  However, in order to maintain the quasi-steady nature of these profiles, the maximum entropy input in a typical heating event cannot be much greater than $10-20 \, {\rm keV \, cm^2}$.  Otherwise the cluster core would linger at much higher entropy levels, because it takes much longer to cool from a high-entropy state, owing to the steeper-than-linear dependence of $t_c$ on $K$.  The dashed line in Figure~\ref{profs+mod+10} shows that simply adding a constant $10 \, {\rm keV \, cm^2}$ entropy pedestal to the pure cooling model produces a reasonable facsimile to the observed profiles out to $\gtrsim 30 \, {\rm kpc}$, another clue to the mode of heat input. 

The structure of these cluster cores from $\sim 10 \, - 100 \, {\rm kpc}$ can be approximated with a simple power-law model that will be useful when we consider heat input.  Figure~1 shows that the entropy profiles in this range are approximately $K(r) \approx 150 \, {\rm keV \, cm^2} \, (r/100 \, {\rm kpc})$, and the temperature gradients at these radii in cooling-flow clusters typically have a power-law slope $d \ln T / d \ln r \approx 1/3$ \citep[e.g.,][]{vf04, don05_cc}.  Combining these facts, we can reverse-engineer the observed density profile $n_e (r) \approx 6.1 \times 10^{-3} \, {\rm cm^{-3}} \, T_{5,100}^{3/2} \, (r/100 \, {\rm kpc})^{-1}$, where $T_{5,100}$ is the gas temperature 100~kpc from the cluster center in units of 5~keV.  The product $r \rho(r) \approx3.7 \times 10^{-3} \, {\rm g \, cm^{-3}} \, T_{5,100}^{3/2}$ is therefore approximately constant in this radial range, and is consistent with the observations of \citet{don05_cc}.

Before we discuss heat input, we have a comment on this power-law model.  Notice that the gas temperature drops by a factor of two from 100 kpc to 10 kpc.  Inside of 10 kpc, the gas temperature 
must then level off, and may even rise a little, if the gas within 10 kpc is isentropic.  Thus, a heating mechanism that maintains an inner entropy plateau at $\sim 10 \, {\rm keV \, cm^2}$ for long periods of time would account for the spectroscopic finding that gas temperatures in the centers of cooling-flow clusters drop to about half the ambient temperature, even in a quasi-static configuration in which there is little net cooling.  

\section{A Framework for Heat Input}

Consider how an energetic outburst from an AGN heats gas with this power-law configuration.  An outflow with a kinetic power output $P_{\rm kin} \sim 10^{45} \, {\rm erg \, s^{-1}}$ shock-heats the innermost portions of the core.  Even if the flow is bipolar, an ellipsoidal cocoon of shocked intracluster gas develops around the outflow, distributing heat through $4 \pi$ steradians \citep[e.g.,][]{Scheuer74, bc89}. If the outflow shuts down after $\sim 10^7$ years, the energy introduced, amounting to a few times $10^{59} \, {\rm erg}$, continues to drive the shock until it becomes subsonic.  The hot bubbles inflated by the outflow then become buoyant, floating away from the center and gradually thermalizing the $P dV$ work they exert on their surroundings \citep[e.g.,][]{Churazov01,Begelman01}.  Here we show that the entropy boosts imparted by a series of such episodic heating events are broadly consistent with the observed entropy profiles of cooling-flow clusters.  

\subsection{Outflow-Driven Shock Heating}

We can estimate the entropy introduced during the outflow-driven stage using the scaling behavior of an outflow-driven bubble and an entropy jump condition.  The shock velocity of such a bubble scales as $v \approx f_P (P_{\rm kin} / \rho r^2)^{1/3}$, where $f_P$ is a structure factor of order unity that depends on the outflow geometry and the preshock density profile \citep[e.g.,][]{om88}.  In order to assess the entropy created by this shock, we apply the approximate entropy jump condition
\begin{equation}
    \Delta K = K_2 - K_1 \approx \frac {\mu m_p v^2} {3(4 n_e)^{2/3}} - 0.16 \, K_1   \; \; ,
\end{equation}
where $K_1$ and $K_2$ are the pre- and post-shock entropy levels, respectively, and $n_e$ is the preshock electron density \citep{vbblb03}.  For the purposes of this estimate, we will ignore the insignificant $0.16 \, K_1$ term.   The entropy increase $\Delta K_P$ induced by such an outflow-driven shock in a medium with a constant value of $r \rho(r)$ is then independent of radius, with $\Delta K_P \propto P_{\rm kin}^{2/3} (r \rho)^{-4/3}$.  Its physical value in the power-law core model of \S\ref{sec-coreprofs} is 
\begin{equation}
   \Delta K_P  \approx  23 \, {\rm keV \, cm^2} \;  \, P_{45}^{2/3} \, T_{5,100}^{-2} \, f_P^2   \; \; ,
\end{equation}
where $P_{45} \equiv P_{\rm kin} / 10^{45} \, {\rm erg \, s^{-1}}$.    

This mode of entropy injection has the very attractive feature of introducing a constant entropy boost throughout the shock-heated zone, as suggested by the observations.  Outflow-driven shocks create no entropy inversions and therefore do not induce the kinds of mixing that would destroy the iron gradients that have built up in cooling-flow clusters.  After the shock passes and the gas settles back into equilibrium, the stratification of the cluster core remains the same because the ordering of specific entropy among gas parcels has not changed.  If this is indeed the primary mode of entropy injection within the centers of clusters, then the observed $\sim 10 \, {\rm keV \, cm^2}$ entropy plateau implies a kinetic power output $\sim 10^{45} \, {\rm erg \, s^{-1}}$ during each heating event.  Such an outflow-driven shock heats a region extending to a radius 
\begin{equation}
  r_P  \approx 16 \, {\rm kpc} \, \left( \frac {\Delta K_P} {10 \, {\rm keV \, cm^2}} \right)^{3/8} \, 
  		T_{5,100}^{3/8} \, t_7^{3/4}
\label{eq-rl}
\end{equation}
from the center, where $t_7 \cdot 10^7 \, {\rm yr}$ is the timescale of the outburst, and the shock's Mach number is ${\cal M} \approx 2.1 (\Delta K / K_1)^{1/2}$, implying that it can remain supersonic to a radius 
\begin{equation}
  r_{\rm ss} \approx 30 \, {\rm kpc} \, \left( \frac {\Delta K_P} {10 \, {\rm keV \, cm^2}}  \right)
\label{eq-rssl}
\end{equation}
in the power-law medium we have assumed.

\subsection{Energy-Driven Shock Heating}

After the kinetic outflow shuts off, the deposited energy, $E \approx (3 \times 10^{59} \,{\rm erg}) P_{45} t_7$, can drive the shock, if the front is still supersonic.  The velocity of an energy-driven shock is $v \approx f_E (E / \rho r^3)^{1/2}$, where $f_E$ is another structure factor analogous to $f_P$, leading  to an entropy boost $\Delta K_E \propto \rho^{-5/3} r^{-3}$.  In the power-law core model, this entropy boost declines with radius:
\begin{eqnarray}
  \Delta K_{\rm E}(r) & \approx & 19 \, {\rm keV \, cm^2} \, 
  				P_{45} t_7 f_E^2 T_{5,100}^{-5/2}  \; \; \; \; \; \; \; \; \; \nonumber \\
 ~ & ~ &   \; \; \; \; \; \; \; \; \; \; \;     \times\left(  \frac {r} {20 \, {\rm kpc}} \right)^{-4/3} \; \; .
\end{eqnarray}
Making the reasonable assumption that $\Delta K_E \approx \Delta K_P (r/r_P )^{-4/3}$, one finds that the energy-driven shock becomes subsonic at
\begin{equation}
  r_{\rm ss} \approx 21 \, {\rm kpc} \, \left( \frac {\Delta K_P} {10 \, {\rm keV \, cm^2}} \right)^{3/14} \,
  		T_{5,100}^{3/14} t_7^{3/7} \; \; .
\label{eq-rsse}
\end{equation}
Notice that this form of entropy injection will cause an entropy inversion if the outburst is too powerful and too brief.  That happens because an energy-driven shock slows down more quickly than an outflow-driven shock, raising the innermost gas to a higher entropy level than gas farther from the center.  Boosting the entropy level in the inner $\sim 30 \,  {\rm kpc}$ of a cluster by a uniform $\sim 10 \, {\rm keV \, cm^2}$ is therefore most naturally done by an outburst lasting several times $10^7$ years.

\subsection{Bubble Heating}

Once the outer shock front becomes subsonic, the plasma bubble inflated by the AGN outflow will buoyantly rise through the intracluster medium.  Inflating the bubble has lifted some of the intracluster medium, increasing its gravitational potential energy by an amount corresponding to the work required to inflate the bubble.  As the bubble subsequently rises, the gas around the bubble falls.  Assuming that the falling gas locally thermalizes that liberated gravitational potential energy, one finds that the bubble introduces heat energy amounting to
\begin{equation}
 \rho V_{\rm bub} \, g \, \Delta r = V_{\rm bub} \left| \frac {dP} {dr} \right| \, \Delta r \; \;  ,
\end{equation}
where $V_{\rm bub}$ is the bubble volume, into each radial interval $\Delta r$ \citep{Churazov01,Begelman01}.  

The amount of additional work a bubble does as it rises and expands, keeping $P V_{\rm bub}^{\gamma_{\rm bub}}$ constant, depends on the polytropic index $\gamma_{\rm bub}$ of its contents.  If the bubble remains coherent as it ascends, it distributes the energy $E_{\rm bub}$ required to create it according to
\begin{equation}
  \frac {dE} {dr} = \xi_{\rm bub} \frac {E_{\rm bub}} {r_{\rm inj}}  
  			\left( \frac {r} {r_{\rm inj}} \right)^{-\xi_{\rm bub}-1} \; \;  ,
\end{equation}
where $r_{\rm inj}$ is the radius at which the bubble is injected and $\xi_{\rm bub} \equiv 2 (\gamma_{\rm bub} - 1) / 3 \gamma_{\rm bub}$, equaling 1/6 if the bubble contents are relativistic and 4/15 if they are non-relativistic. Here we will assume that bubbles are injected at the radius $r_{\rm ss}$ where the shock driven by the outburst becomes subsonic.  Those rising bubbles can generate additional entropy by causing turbulence that mixes low-entropy gas from the cluster core with the higher-entropy gas lying above it.  However, we will not consider mixing processes here because they  do not increase the total amount of thermal energy in the intracluster medium.

Heating the gas by an amount $\Delta q$ per particle augments its specific entropy by $\Delta \ln K^{3/2} = \Delta q / kT$, implying $\Delta K = (2/3 n_e^{2/3}) \Delta q$.  The total amount of bubble heating per particle from a single outburst is 
\begin{equation}
  \Delta q (r) = \frac {\mu m_p} {4 \pi r^2 \rho} \frac {dE} {dr} \; \;  ,
\end{equation}
implying that the entropy boost in the bubble-heated zone is $\Delta K_{\rm bub} \propto \rho^{-5/3} r^{-3-\xi_{\rm bub}}$.  Defining $f_{\rm bub} = E_{\rm bub} / E$ to be the fraction of the input energy left in bubble form when the shock becomes subsonic, we obtain the following expression for bubble heating as a function of radius in the power-law core model
\begin{eqnarray}
  \Delta K_{\rm bub}(r) & \approx & 4.0 \, {\rm keV \, cm^2} \, 
  			\xi_{\rm bub} f_{\rm bub} f_E^{-2} \; \; \; \;  \; \; \; \; \;  \; \;  \; \; \; \;  \; \; \;  \; \; \; \; \nonumber \\ 
      ~ & ~ & \; \; \;  \; \; \times
  	\left(  \frac {r_{\rm ss}} {30 \, {\rm kpc}} \right)  
	\left(  \frac {r} {r_{\rm ss}} \right)^{- \frac {4} {3} -\xi_{\rm bub} }.  
\end{eqnarray}
Like an energy-driven shock, this form of heating will also cause an entropy inversion if too much bubble energy is injected at too small a radius because its entropy injection profile falls even more rapidly with radius than the profile for the energy-driven shock.

\subsection{Net Entropy Change}

A single AGN outburst thus generically heats the cluster core in three zones, an inner outflow-driven zone, an intermediate energy-driven zone, and an outer bubble-heated zone.  Disturbances moving at the sound speed propagate to 100~kpc in $\sim 10^8$ years, during which time the inner regions relax to a new equilibrium configuration with slightly elevated entropy and resume cooling.  In order to evaluate the net entropy change over a single cycle of heating and cooling, we will first determine the entropy losses due to radiative cooling during this time interval and then compare those losses to the entropy boost from a single outburst.

A radiative loss rate of $\dot{q} = n_{\rm i} n_e \Lambda(T) / (n_{\rm i} + n_e)$ carries entropy away from a plasma of ion density $n_i$ at the rate $\dot{K} = - (2/3)[n_{\rm i}/(n_{\rm i}+n_e)] n_e^{1/3} \Lambda (T)$.  Pure free-free cooling therefore removes an entropy
\begin{eqnarray}
  \Delta K_c (r) & \approx &  - \, 9.3 \, {\rm keV \, cm^2} \; T_{5,100}  \; \; \; \; \; \; \; \; \nonumber \\
 ~ & ~ & \; \; \; \; \;  \times \left( \frac {r} {10 \, {\rm kpc}} \right)^{-1/6} \frac {\Delta t} {10^8 \,{\rm yr}} \; \;
\end{eqnarray}
during a time interval $\Delta t$, given the simple power-law model we have been using for the core gas configuration.  Notice that if outflow-driven shock heating can compensate for cooling at small radii, then heating and cooling will remain nearly balanced out to $r_P $ because of the weak dependence of  $\Delta K_c$ on radius.  However, in the power-law core model, cooling will dominate heating at $r \gg r_P $ over the course of one cycle because both energy-driven shock heating and bubble heating fall off more quickly with radius than cooling.

\begin{figure}[t]
\includegraphics[width = 3in,trim = 1.0in 1.0in 0.8in 1.0in, clip]{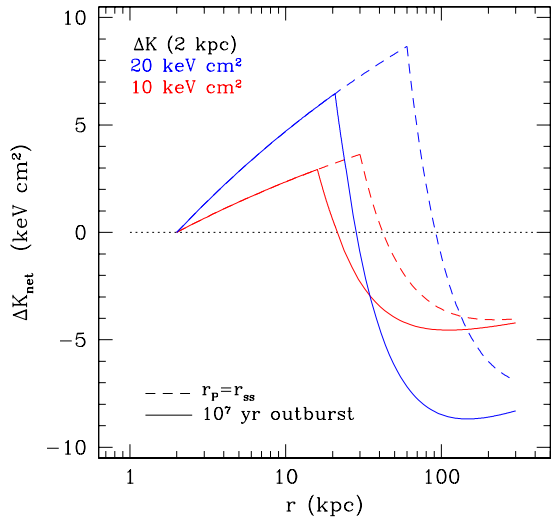}
\caption{ \footnotesize
Net entropy change $\Delta K_{\rm net}$ during a single cycle of heating and cooling given the power-law model of core structure.  The upper set of lines (blue) shows models in which each AGN outburst heats the intracluster medium at a radius of 2~kpc by 20~keV~cm$^2$.  The lower set of lines (red) shows models in which each AGN outburst heats the intracluster medium at a radius of 2~kpc by 10~keV~cm$^2$.  The cooling time after each outburst is normalized to give zero net entropy change at that radius.  Solid lines show the net entropy change for an outburst lasting $10^7$~years that deposits heat according to the piecewise power law described in the text.  Dashed lines show the net entropy change for outbursts lasting long enough for the heating to remain outflow-driven until the shock front becomes subsonic.  The maximum net entropy gain occurs at the radius $r_P $ the shock has reached at the time the outburst ceases.  Only the longer outbursts are able to prevent net cooling at $\gtrsim 30$~kpc, suggesting that each AGN outburst must remain at $\sim 10^{45} \, {\rm erg \, s^{-1}}$ for several times $10^7$~years in order to shut off the cooling flow while maintaining a quasi-steady inner entropy profile.
\label{delk_pl}}
\end{figure}

Thus, in order for heating to compensate for cooling throughout the $\sim 100 \, {\rm kpc}$ region where the cooling time is less than a Hubble time, the outflow-driven zone must be relatively large, implying that outbursts capable of stopping a cooling flow indeed last several times $10^7$ years.  Figure~\ref{delk_pl} illustrates this idea in the context of the power-law core model.  Solid lines show the net entropy change over the course of a single cycle of heating and cooling for outbursts lasting $10^7$ years.  Two different cases are shown, corresponding to entropy boosts of $\Delta K_P = 10 \, {\rm keV \, cm^2}$ and $20 \, {\rm keV \, cm^2}$ in the outflow-driven zone.  Outside of that, in the energy-driven zone, the boost in each case is $\Delta K_P (r/r_P )^{-4/3}$, where $r_P $ is given by equation~(\ref{eq-rl}) with $T_{5,100} = 1$.  This zone extends to $r_{\rm ss}$ as given by equation~(\ref{eq-rsse}), and beyond that point bubble heating is assumed to raise the entropy by $\Delta K_P (r_{\rm ss}/r_P )^{-4/3} (r/r_{\rm ss})^{-3/2}$, so that the entropy jump is continuous across $r_{\rm ss}$.  The validity of that assumption depends on the structure factors $f_P$, $f_E$, and $f_{\rm bub}$ and needs to be checked with numerical simulations.  The power-law index in radius here applies to the relativistic bubble case, but the non-relativistic case is virtually the same, with an $r^{-1.6}$ dependence.

Because the time interval between heating episodes depends on the entropy boost at small radii, we choose $\Delta t$ so that heating matches radiative losses at a radius of 2~kpc.  Outside of that radius, the net entropy gain is several keV~cm$^2$ out to about $r_P $, where there is a sharp turnaround.  Beyond about 30~kpc the net entropy change is negative, even for a central entropy boost of 20~keV~cm$^2$.  The dashed lines show how the picture changes when an outburst lasts long enough for $r_P $ to equal $r_{\rm ss}$ as given by equation~(\ref{eq-rssl}).  Then the outflow-driven zone extends much farther, and in the case of a 20 keV~cm$^2$ boost, heating can exceed cooling out to $\sim 100$~kpc.  The time required for an outflow-driven shock to reach $r_{\rm ss}$ is $2.4 \times 10^7 \, T_{5,100}^{-1/2} \, {\rm yr}$ in the 10 keV~cm$^2$ case and  $4.3 \times 10^7 \, T_{5,100}^{-1/2} \, {\rm yr}$ in the 20 keV~cm$^2$ case.

\begin{figure}[t]
\includegraphics[width = 3in,trim = 1.0in 1.0in 0.8in 1.0in, clip]{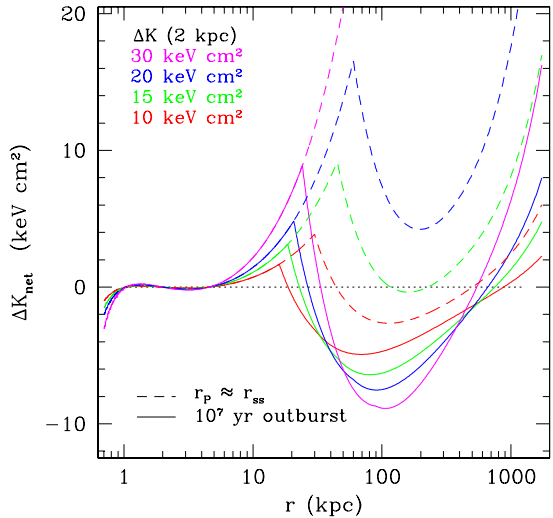}
\caption{ \footnotesize
Net entropy change $\Delta K_{\rm net}$ during a single cycle of heating and cooling when the pure cooling model of Voit et al. (2002) is used for the pre-outburst structure of the core.   Each set of lines shows two models with  the same value of entropy input at 2~kpc.  From top to bottom, these values are 30~keV~cm$^2$ (magenta), 20~keV~cm$^2$ (blue), 15~keV~cm$^2$ (green), and 10~keV~cm$^2$ (red).   Cooling after each outburst is normalized to give zero net entropy change at 2~kpc.  Solid lines show the net entropy change for an outburst lasting $10^7$~years.  Dashed lines show the net change for outbursts lasting long enough for the heating to remain outflow-driven until the shock front becomes subsonic.  Interestingly, an outburst that raises the central entropy level to 15~keV~cm$^2$ marginally compensates for cooling at all radii over the course of a cycle, suggesting that the characteristic minimum central entropy level of $\sim 10 \, {\rm keV \, cm^2}$ is determined by the minimum outburst required to halt a cooling flow.
\label{delk_mod78}}
\end{figure}

A slightly more refined picture emerges when we take the core structure at the time of the outburst to be that of the pure cooling model of Voit et al. (2002), corresponding to the solid line in Figure~\ref{profs+mod+10}.  In that model of the core, gas density starts to drop with radius more rapidly than $r^{-1}$ outside of $\sim 30$~kpc, gradually increasing the efficiency with which all forms of heating increase the entropy.  Figure~\ref{delk_mod78} shows models in which the entropy boost from the outburst is a continuous piecewise power law that is $\propto (r \rho)^{-4/3}$ in the outflow-driven zone, $\propto \rho^{-5/3} r^{-3}$ in the energy-driven zone, and $\propto \rho^{-5/3} r^{-3-\xi_{\rm bub}}$ in the bubble-heated zone.  Boundaries between the zones are the same as those used for Figure~\ref{delk_pl}.  The entropy boost is normalized so that $\Delta K_P$ at 2~kpc takes values of 10, 15, 20, and 30 ~keV~cm$^2$, and cooling is assumed to match heating at that radius.  Solid lines show models for outbursts lasting $10^7$ years, and dashed lines show models for outbursts long enough for the outflow-driven zone to reach the radius at which the shock becomes subsonic. 

The simple models illustrated in Figure~\ref{delk_mod78} suggest that episodic outbursts that impart $\sim 15 \,{\rm keV \, cm^2}$ to the central regions of a cluster and last until the shock reaches $\sim 30$~kpc can compensate for radiative cooling throughout an entire cluster.  In all the models, heating balances cooling from 1~kpc to 5~kpc with a high precision that is likely to be coincidental, given the crudity of the model.  Outside of 5~kpc, the net entropy gain per cycle rises out to the outer radius of the outflow-driven zone, beyond which the net entropy gain declines because the shock decelerates more quickly there.  The sharp cusp appearing at the transition to the energy-driven zone is an artifact of the piecewise power-law model for heating; a more realistic model would have a more gradual turnover.  Beyond $\sim 50-100 \, {\rm kpc}$, in the bubble-heated zone, the net entropy change climbs again with radius because the gas density profile gradually steepens there, approaching $\rho \propto r^{-2}$ beyond the cluster core.  The entropy gain outside the core from bubble heating is therefore $\Delta K_{\rm bub} \propto r^{1/3 - \xi_{\rm bub}}$, which is nearly constant with radius.  However, bubble heating operates out there only if the bubbles can remain coherent over many pressure scale heights, requiring some sort of stabilizing mechanism to ward off hydrodynamic shredding \citep[e.g.,][]{bk01}.

One intriguing possibility suggested by Figure~\ref{delk_mod78} is that the characteristic minimum entropy level of $\sim 10 \, {\rm keV \, cm^2}$ in cluster cores is determined by the minimum AGN outburst required to halt a cooling flow.  All the cluster entropy profiles in Figure~\ref{profs+mod+10} flatten at roughly this level even though the clusters range over a factor more than three in temperature.  Our analysis indicates that AGN outbursts unable to heat the central regions to this level are also not powerful enough to push outflow-driven shocks to $\sim 30$~kpc, the necessary inner radius for either energy-driven shocks or bubble heating to compensate for radiative cooling in the outer parts of a cluster's core ($\sim 50-100$~kpc).  This feature of our simple model suggests that the AGN feedback mechanism may have become tuned over time so that cooling is kept marginal \citep[see][for an example of such a mechanism]{ob04}.  

An alternative possibility is that the consistency among the inner entropy levels in the sample of Donahue et al. (2005) is a selection effect, because clusters with the lowest central entropy levels will appear to have the strongest cooling flows.  It could be that outbursts of kinetic energy from a cluster's central AGN outbursts vary greatly in power, sometimes imparting entropy boosts several times greater than $10 \,{\rm keV \, cm^2}$.  In that case, however, one would expect a large fraction of clusters to have elevated central entropy levels because it would take much longer for radiative cooling to reduce the central entropy in those clusters.

\subsection{Multiple Outbursts}

Now we will briefly consider the evolution of the intracluster medium over time in the context of this simple framework for AGN heating.  In general, one would not expect precise balance between AGN heating and radiative cooling throughout the cluster core because the conditions governing accretion at $\lesssim 1$~kpc are not closely linked to those at $\sim 100$~kpc.  A common feature of the outburst models presented here is that heating exceeds cooling in the outer parts of the outflow-driven zone, and that excess can persist well beyond the outflow-driven zone if bubble heating is effective.  One therefore expects entropy to accumulate over time in the $\sim 30-100$~kpc range, if the outbursts are strong enough to prevent net cooling everywhere outside of $\sim 1$~kpc.

Evidence for such an accumulation of entropy is present in Figure~\ref{profs+mod+10}.  The entropy profiles shown there tend to exceed the pure-cooling model by $\sim 50 \, {\rm keV \, cm^2}$ at $\sim 100$~kpc.  The models we have constructed here are too crude for a detailed comparison with that excess to be fruitful.  Statistical variations in the duration and kinetic power of the outbursts will cause the size of the outflow-driven zone to differ from outburst to outburst, and the shape of that zone is likely to be elongated, owing to the asymmetry of the outbursts.  Neither of those features are represented in our spherically-symmetric, single-outburst models.

Nevertheless, a generic prediction of this framework is that highly energetic outbursts should limit the total accumulation of entropy at large radii over the course of several billion years.
From equation~(\ref{eq-ctime}) for the cooling time from a given entropy level, one infers that the maximum number of outbursts that can occur over a time period $t$ is
\begin{eqnarray}
   N_{\rm bursts} & \sim & 50 \, \left( \frac {\Delta K_P} {10 \, {\rm keV \, cm^2}} \right)^{-3/2} 
   						 \left( \frac {kT} {5 \, \keV} \right)  \nonumber \hspace{2em} \\
	~ & ~ & \hspace{6em} \times  \left( \frac {t} {5 \, {\rm Gyr}} \right) \; \; ,
\label{eq-nbursts}
\end{eqnarray}
if each outburst imparts an entropy boost $\Delta K_P$ in the outflow-driven zone.  Thus, several tens of $\sim 10 \, {\rm keV \, cm^2}$ outbursts can occur, but only a few at the $\sim 50 \, {\rm keV \, cm^2}$ level.  If the excess entropy introduced at large radii by a single outburst is $\propto \Delta K_P$, then the accumulated excess entropy arising from a series of such outbursts is $\propto \Delta K_P^{-1/2}$.  However, an adequate calibration of the relation between the central entropy input from an outburst and the corresponding excess entropy it produces will require high-resolution numerical simulations.

In principle, it may be possible to use the excess entropy observed in the outer regions of clusters to constrain the outburst history.  Figure~\ref{delk_mod78} indicates that each outburst incrementally deposits several keV~cm$^2$ at $\sim$1~Mpc, where cooling is negligible, as long as bubble heating still operates at such large radii.  For example, suppose a series of $\sim 15 \, {\rm keV \, cm^2}$ outbursts were to introduce $\sim 5 \, {\rm keV \, cm^2}$ at $\sim 1$~Mpc with each outburst, as suggested by Figure~\ref{delk_mod78}.  Roughly thirty such outbursts could occur in a $\sim 5$~keV cluster over a period of 5~Gyr, according to equation~(\ref{eq-nbursts}), leading to an accumulated excess entropy $\sim 150 \, {\rm keV \, cm^2}$ at large radii.  The entropy excesses seen in the outer parts of $\lesssim 5 \, {\rm keV}$ clusters \citep{psf03} might therefore result from the cumulative effects of bubble heating during the cluster's lifetime \citep[see also][]{rrnb04}.  

In even smaller systems with temperatures $\lesssim 1$~keV, excess entropy that accumulates at $\gtrsim 100$~kpc may come to dominate the gravitationally produced entropy.  Because these systems have smaller core radii than clusters, AGN outbursts can penetrate more easily to radii beyond $\sim 30$~kpc, depositing more entropy there.  Thus, the ``entropy deficit'' observed in the outskirts of groups by \citep{Mahdavi05} might not be an actual deficit in the gravitationally produced entropy but rather just a deficit with respect to a large entropy excess of non-gravitational origin that has built up near the core radius.

\subsection{Outbursts of Sustained Power}

The dramatic heating event observed in MS0735 +7421 by \citep{McNamara05} inspires us to add one more wrinkle to this framework for AGN heating.  So far, we have considered only outbursts of $\sim 10^{45} \, {\rm erg \, s^{-1}}$ lasting significantly less than $10^8$ years, but MS0735+7421 appears to have been sustaining a kinetic power output $\sim 1.7 \times 10^{46} \, {\rm erg \, s^{-1}}$ for $\sim 10^8$~years.  The observed central entropy level of $\sim 30 \, {\rm keV \, cm^2}$ is consistent with a constant-power outburst at this level, given a structure factor $f_P \sim 0.5$, and corresponds to a central cooling time $\sim 1$~Gyr at the central temperature of 3~keV.  In those respects MS0735+7421 is similar to the two radio-quiet cooling-core clusters observed by \citet{don05_rq}, which also have $\sim 30 - 50 \, {\rm keV \, cm^2}$ entropy levels and $\sim 1$~Gyr central cooling times, suggesting that these clusters are what MS0735+7421 will look like a few hundred million years from now.  If that is the case, then instances of powerful, sustained outflows from AGNs in clusters might not be all that rare and should be incorporated into the overall framework for AGN heating.

How much longer the energetic outflow in MS0735+7421 will persist is anyone's guess, but its effects on the cluster's overall entropy profile are likely to be profound and lasting.  The outflow has already propagated to a radius $\sim 300$~kpc, leaving the cluster core and entering the region where gas density is steepening to a $\rho \propto r^{-2}$ density profile.  An interesting feature of an outflow-driven shock in a medium with that density distribution is that it propagates at constant velocity, generating an entropy boost that is proportional to the original entropy, i.e. $\Delta K / K$ is constant.  A sustained, high-power outflow therefore raises the outer entropy profile of a cluster by a constant factor, preserving its shape.

That sort of entropy boost is particularly interesting in light of {\em XMM-Newton} observations showing that the outer entropy profiles of low-temperature clusters have shapes similar to those of hotter clusters but have unexpectedly large normalizations, indicating that some non-gravitational process has elevated the entropy by a constant factor throughout the intracluster medium  \citep{pa03, pa05}.  One suggested possibility for causing this global entropy boost is that early galactic winds smoothed the intergalactic medium of the protocluster before the cluster formed, thereby amplifying the amount of entropy generated in the cluster's accretion shocks \citep{vbblb03, psf03}.  However, simulations of entropy amplification suggest that supernova-driven winds have trouble generating enough smoothing to account for the observed entropy excesses \citep{Borg05}.  Smoothing by AGNs might be required, or maybe it is possible for the central AGN to produce the entire entropy excess through large outbursts alone.

In order to assess the large-scale impact of heating events like that in MS0735+7421, we can consider an idealized cluster model in which the cluster atmosphere has $\rho \propto r^{-2}$ everywhere.  Taking the baryon fraction to equal the global value gives $\rho r^{-2} \approx 3.2 \times 10^{21} \, {\rm g \, cm^{-1}} \, (kT / 5 \, {\rm keV})$, where $T$ is now the mean temperature of the cluster.  The entropy boost produced in that density profile by a sustained outburst of kinetic power output is
\begin{eqnarray}
	\frac {\Delta K} {K} & \approx &  \frac {\mu m_p f_P^2} {3 \cdot 4^{2/3} kT}
						\left( \frac {P_{\rm kin}} {\rho r^2} \right)^{2/3} \nonumber \\
	~ & \approx & 0.35 \, f_P^2 P_{46}^{2/3} \left( \frac {kT} {5 \, {\rm keV}} \right)^{-5/3} \; \; ,
\end{eqnarray}
where $P_{46} \equiv P_{\rm kin} / 10^{46} \, {\rm erg \, s^{-1}}$.  In other words, the outburst event in MS0735+7421 is destined to raise its global entropy profile by $\gtrsim 10$\%, depending on the structure factor $f_P$.  Furthermore, lower-temperature clusters are subject to even larger global boosts, implying that strong, sustained outbursts of AGN heating could be a substantial contributor, maybe even the dominant contributor, to the entropy excesses observed at large radii in low-temperature clusters.

\section{Summary}

In this paper we have outlined a framework for AGN heating of cluster cores inspired by the observed core entropy profiles of cooling-flow clusters  \citep{don05_cc}.   We have focused primarily on entropy rather than on density or temperature, because entropy is the quantity most directly affected by heat input and radiative losses.  Tracking entropy changes allows one to predict, at least qualitatively, how an outburst of AGN energy becomes distributed with radius in the intracluster medium.   Applying these simple predictions to the observed distributions of entropy, temperature, and density in cluster cores indicates how the entropy at each radius should change over a single cycle of outburst heating and radiative cooling.  The picture we arrive at from considering the time-dependent behavior of these core entropy profiles has much in common with other AGN-heating models that were inspired by other aspects of clusters with cool cores, especially the X-ray cavities often found there.   Here we summarize the AGN-heating framework that emerges from considerations of intracluster entropy. 

The key features of the core entropy profiles in cooling-flow clusters, described in Section~2, are as follows.
\begin{enumerate}
\item The profiles observed by \citep{don05_cc} are very similar to one another when plotted as functions of radius in unscaled physical units, even though the clusters range over a more than a factor of three in temperature.  Power-law fits to these profiles show that $K \propto r^\alpha$ with $\alpha \approx 0.9-1.3$, and the normalizations of the profiles are consistent with earlier indications that entropy at a fixed fraction of the virial radius is $\propto T_\phi^{2/3}$, where $T_\phi$ is the characteristic temperature of the cluster's halo \citep{psf03,vp03, pa03, Piffaretti05}.   A power-law model with $K(r) \approx 150 \, {\rm keV \, cm^2} \, (r/100 \, {\rm kpc})$, $T \propto r^{1/3}$, and $\rho \propto r^{-1}$ is a reasonable approximation to the gas configuration from $\sim 10$~kpc to $\sim 100$~kpc.  
\item In the outer regions of these cluster cores, the entropy profiles appear to approach the pure-cooling cluster model of \citet{vbbb02}, in which a cluster whose initial configuration is determined by hierarchical structure formation is allowed to cool for a Hubble time without feedback.  The pure-cooling model is apparently a bounding lower envelope to the allowed set of entropy profiles.
\item The inner regions of these profiles flatten out at an entropy level $\sim 10 \, {\rm keV \, cm^2}$.  Furthermore, simply adding an entropy pedestal of $10 \, {\rm keV \, cm^2}$  to the pure-cooling model at all radii reproduces the observed shape of these profiles at radii from $\sim 1$~kpc to $\gtrsim 30$~kpc.
\item  This inner entropy minimum corresponds to a cooling time $\sim 10^8$~years, suggesting that the heating mechanism producing the entropy pedestal is episodic on roughly this timescale \citep[see also][]{David01,kb03}.  If AGN heating is indeed episodic, then the relatively strong power-law dependence of cooling time on entropy implies that the central entropy level remains near its maximum value during most of the time period between feedback episodes.  Thus, typical AGN heating events probably impart $\sim 10-15 \, {\rm keV \, cm^2}$ in the central parts of the core.
\item Because the entropy pedestal introduced by a heating episode persists over most of the cycle, the central temperature gradient should remain approximately flat within $\sim 10$~kpc, which corresponds to a temperature that is approximately one-half the value at $\sim 100$~kpc, given the $T \propto r^{1/3}$ temperature gradient outside of 10~kpc \citep[see also][]{kb03}.  In this configuration, the lower bound to the gas temperatures observed in cooling-flow clusters arises in a quasi-steady hydrostatic configuration with an inner entropy floor and reflects the potential-well depth in the region where the entropy distribution flattens. 
\end{enumerate}
In addition to these features of the entropy profiles, the strong iron abundance gradients seen in the cores of cooling-flow clusters provide another important constraint on the heating mechanism:  it cannot induce too much mixing in the core.  Otherwise, it would wipe out the observed gradients \citep[e.g.,][]{bmcic02}.

Taken as a whole, these features of the entropy structure in cluster cores point toward a specific framework for AGN heating, independent of observations of X-ray cavities (see \S3).
\begin{enumerate}
\item An AGN outburst with a constant kinetic power output naturally produces a constant entropy pedestal if it propagates into a medium with $\rho \propto r^{-1}$.   Raising the inner entropy by $\sim 10 \, {\rm keV \, cm^2}$ requires outbursts of $\sim 10^{45} \,{\rm erg \, s^{-1}}$, given the values of $r \rho (r)$ observed in the cores of cooling-flow clusters.  This mode of heating is particularly attractive because a boost in entropy that is constant with radius does not produce entropy inversions, which are not generally observed and would induce convection, mixing, and reduction of the central iron gradient.
\item In order to produce an entropy pedestal extending to $\sim 30$~kpc, as suggested by the observed profiles, an outburst sufficient to raise the central entropy by $\sim 10 \, {\rm keV \, cm^2}$ needs to last several times $10^7$ years.  Shorter bursts pump too much outburst energy into too small a volume, leading to entropy inversions, because both energy-driven shocks and bubble heating produce entropy boosts that decline with radius in a medium where $\rho \propto r^{-1}$.   
\item The leading shock of an outburst that raises the central entropy by $\sim 10 \, {\rm keV \, cm^2}$ and lasts for $\sim 3 \times 10^7$~ years will become subsonic and decay to sound waves in the neighborhood of $\sim 30$~kpc from the center.  At that point the hot bubble of ejected material that was driving the shock can become buoyant.  If the intracluster medium efficiently thermalizes the work done by that bubble as it rises, then it can offset cooling at large radii in the core \citep{Churazov00,Churazov01,Churazov02}.  Small bubbles can be just as effective as large ones, as long as the total amount of initial bubble energy is the same \citep{Begelman01,rb02}.  One can therefore compute the entropy injected at all radii by an outburst, given the size of the central entropy boost and the duration of the outburst, under the assumption that the bubbles remain coherent as they rise.
\item Assuming that the outbursts are regulated by the need for heating to balance radiative cooling at small radii over the course of an outburst cycle determines the net entropy change at all radii from a single cycle.  The size of the central entropy boost determines the number of outburst cycles that can occur during the life of a cluster, and therefore regulates the accumulation of excess entropy at larger radii.  
\item In order for bubble heating to exceed cooling in the outer parts of the core over the course of a cycle, the inner radius of the bubble-heated region must be $\gtrsim 30$~kpc.  That requirement places a lower limit $\sim 10-15  \, {\rm keV \, cm^2}$ on the size of the central entropy boosts.  Weaker outflows do not extend the base of the bubble-heated zone to a sufficiently large radius.  This feature of cluster cores may explain the ubiquity of the $\sim 10 \, {\rm keV \, cm^2}$ pedestal.  Perhaps the feedback mechanism has become tuned over time to produce outbursts with power output and duration sufficient to ward off cooling catastrophes.
 \item Outside the cluster core, the gas density profile steepens to $\rho \propto r^{-2}$, thereby increasing the efficiency of entropy production.  Sustained outbursts with constant kinetic power output can therefore produce entropy boosts for which $\Delta K/K$ is constant with radius, raising the previous entropy profile by a constant factor without changing its shape.  An outburst power of $\sim 10^{46} \, {\rm erg \, s^{-1}}$, such as that observed in MS0735+7421, can raise the outer entropy profile of a 5~keV cluster by $\gtrsim 10$\%, and the boosts in lower-temperature clusters can be substantially greater.  Strong AGN outbursts can therefore potentially account for the elevated outer entropy profiles of low-temperature clusters \citep[see also][]{rrnb04}.
\end{enumerate}

Interestingly, the picture that emerges from consideration of the core entropy profiles alone calls for outbursts that deposit a total energy $\sim 10^{60} \, {\rm erg}$ every $\sim 10^8$~years, corresponding to a time-averaged AGN power output of several times $10^{44} \, {\rm erg \, s^{-1}}$ with a duty cycle $\sim 20-30$\%.  Dividing this energy between two bubbles, each requiring an energy of $4PV$ to produce, implies that $PV \sim 10^{59} \,{\rm erg}$ in each bubble.  All of these characteristics are remarkably consistent with the properties of AGN outbursts inferred from X-ray cavities \citep[e.g.,][]{brmwn04}

Further progress within this framework will require numerical simulations to determine the actual entropy boost imparted by a single outburst as a function of radius, given an initial configuration for the core.  There has already been a substantial amount of work in this area, ranging over a wide volume of parameter space in kinetic power, outburst duration, and core configuration at the time of the outburst \citep[recent papers include][]{DallaVec04, obbs04, Zanni05}.   The analysis we have presented indicates the most fruitful area of this parameter space to explore in future work, while strengthening the case for episodic AGN outbursts as the mechanism that limits cooling and condensation in cooling-flow clusters at the present time and therefore the growth of the largest galaxies in the universe.  A complete picture of the intracluster medium will also need to account for earlier processing of a cluster's intergalactic gas by cooling and supernova feedback during the epoch of galaxy formation, whose effects can be amplified by accretion shocks as the cluster forms \citep{vbblb03, Borg05}.  

\vspace*{1em}
We thank B. McNamara, G. Bryan, and A. Babul for their comments on an earlier draft of this paper.  We also thank the NASA Astrophysics Theory Program for supporting this work under grant number NNG04GI89G.


\end{document}